\documentclass[preprint,english,showpacs,floatfix]{revtex4}
\usepackage{babel,amsmath,amssymb,dcolumn,graphicx}
\begin{document}


%
%

\title{VACUUM POLARIZATION EFFECTS ON QUASINORMAL MODES IN ELECTRICALLY
CHARGED BLACK HOLE SPACETIMES}

\author{OWEN PAVEL FERNANDEZ PIEDRA}

\affiliation{Departamento de F\'{i}sica y Qu\'{i}mica, Universidad de Cienfuegos, Carretera a Rodas, Cuatro Caminos, s/n. Cienfuegos, Cuba\\
opavel@ucf.edu.cu}

\author{JEFERSON DE OLIVEIRA}

\affiliation{Instituto de F\'{\i}sica, Universidade de S\~ao Paulo,
  CP 66318,
05315-970, S\~ao Paulo, Brazil\\
jeferson@fma.if.usp.br}


\begin{abstract}
We investigate the influence of vacuum polarization of quantum
massive fields on the scalar sector of quasinormal modes in
spherically symmetric black holes. We consider the evolution of a
massless scalar field on the spacetime corresponding to a charged
semiclassical black hole, consisting of the quantum corrected
geometry of a Reissner-Nordstr\"om black hole dressed by a quantum
massive scalar field in the large mass limit. Using a sixth order
WKB approach we find a shift in the quasinormal mode frequencies due
to vacuum polarization .
\end{abstract}

\pacs{black holes; quasinormal modes; semiclassical gravity.}

\maketitle
\section{Introduction}

The evolution of a small perturbation in a black hole background
geometry gives rise, under appropriate boundary conditions to a
discrete set of complex frequencies called quasinormal frequencies.
Actually, the evolution of a perturbation, taking for instance the
Schwarzschild black hole, can be divided in three stages. The first
stage is a signal highly dependent on the initial conditions.
Intermediate times are characterized by a exponential decay, with
the frequencies determined by the quasinormal modes, that depends
upon the parameters of the black hole. Late times are generally
characterized by a power-law tail \cite{evl}.

Studying the quasinormal spectrum of black holes we can gain  some
valuable information about these objects, since quasinormal
evolution depends only upon the parameters of black hole itself.
Thus these frequencies represents the characteristic resonance
spectrum of a black hole response \cite{kokkotas1}. In addition, we
can investigate the black holes stability against small
perturbations. In this context, several numerical methods have been
developed that allow us to make the necessary calculations with an
appropriate precision. \cite{price,carlemoskono,abdwangetc}.

Further contexts include astrophysical black holes \cite{kokkotas1}
\cite{nollert} and the AdS/CFT correspondence, where the inverse of
imaginary part of quasinormal frequencies of AdS black holes can be
interpreted as the dual CFT relaxation time \cite{horowitz}
\cite{Miranda}.

An interesting problem consists in determining what changes appears
in the quasinormal mode spectrum of a black hole if we consider such
a system surrounded by a quantum field with a semiclassical gravity,
leading to a quantum corrected line element for the dressed black
hole. The stress energy tensor of the quantum field surrounding the
classical compact object contains all the necessary information to
treat the above mentioned problem, as it enters as a source in the
semiclassical Einstein equations leading to a dressed black hole
solution \cite{birrel}. Unfortunately, the exact determination of
the quantum stress energy tensor in the general case is a very
difficult task, and consequently there exist in the literature
several approaches to obtain this quantity, including numerical
ones. Due to the fact that the principal interest in further
applications is not the quantum stress energy tensor itself, but
rather its functional dependence on a wide class of metrics, it is
obvious that we need, at least partially, to use approximate methods
to develop a tractable expression for this quantity. Also there
exist another difficulty with semiclassical gravity, and is the fact
that the quantum fluctuations of the metric and the associated
additional graviton contribution to the renormalized quantum stress
energy tensor are ignored. In fact, the effect of linear gravitons
contribute a term to the stress energy tensor of the same order as
those coming from ordinary matter and radiation fields, and one
should include this contribution. A popular solution to this
difficulty is to study the separated effects of different sorts of
quantized fields in classical spacetimes \cite{birrel,AHS}. In this
way one develops knowledge of the type and range of physical effects
created by quantum fields. If one has examined a quite wide
selection of other quantum fields, then it is justifiable to assume
that the graviton contribution will not be extremely different
\cite{grishuk,AHS}. On the other hand, it is reasonable to ignore
the graviton contribution to the stress energy tensor when one is
computing the metric backreaction caused by other quantized matter
fields alone to first order. A consequence of this approach is that
the backreaction is only meaningful in a perturbative sense, and the
effect of the stress energy tensor for the quantum matter fields is
regarded as a perturbation of the classical spacetime geometry.
Thus, in semiclassical gravity, having a general analytic expression
for the quantum stress tensor, we can find a perturbative solution
to the backreaction problem and seek for a quantum corrected metric.
There exist some results in the literature, since the pioneering
work of York \cite{york}, who solved the semiclassical Einstein
equations for a Schwarzschild black hole dressed by a massless
conformally coupled scalar field, using for the quantum stress
energy tensor the results given earlier by Page \cite{page}.

In a previous paper, Konoplya\cite{konoplyabtz} studied for the
first time the influence of the semiclassical backreaction upon
black hole quasinormal modes. He investigated the particular case of
a BTZ black hole dressed by a massless conformal scalar field, in
which the particle creation around the event horizon dominates over
the vacuum polarization effect. In this particular case, the
2+1-dimensional character of the metric ensures that the quantum
corrected solution is self-consistent in the sense that the only
quantum corrections to the geometry are coming from quantum matter
fields, and there are no corrections from graviton loops.

To make a similar study in the four-dimensional case is more
complicated, for the difficulty mentioned before regarding the
ignorance of graviton contribution to the quantum stress energy
tensor. In addition, for massless fields, the quantum corrected
metric components diverge as \(r\rightarrow\infty\) and to obtain
the correct solutions to the backreaction problem we need to impose
some sort of boundary to the system under study. For the study of
the corresponding quasinormal modes we have to deal with an
effective potential with a step function or a delta function at the
location of the boundary shell. This feature in the effective
potential changes the spectrum. For this reason, we need to develop
a model independent approach for the determination of the
quasinormal frequencies in those cases, in such a way that the
effective potential does not depend on the location of the boundary
needed to fix the semiclassical solution \cite{konoplyabtz}.

However, massive fields are good candidates for investigating the
influence of their quantization on quasinormal modes. Vacuum
polarization effects for very massive fields are not difficult to
compute since there exist an exact analytic approximation for the
quantum stress energy tensor, based on the Schwinger-DeWitt
expansion of the quantum effective action
\cite{owen1,owen2,owenarxiv,matyjasek,avramidi,frolov-bv}. Using the
first non divergent term in this expansion as the effective action
(one-loop approximation), we can obtain a general analytical
expression for the quantum stress energy tensor by functional
differentiation with respect to the metric tensor, and solve the
backreaction problem in a general analytical way. With respect to
the validity of the Schwinger-DeWitt approximation, it is well-known
that this method can be used to investigate effects like the vacuum
polarization of massive fields in curved backgrounds, whenever the
Compton's wavelenght of the field is less than the characteristic
radius of curvature
\cite{frolov,barvinsky,DeWitt,avramidi,matyjasek}.

This paper is devoted to find this vacuum polarization effects upon
quasinormal modes of quantum corrected Reissner-Nordstr\"om black
holes in four dimensions. To the best of our knowledge this is the
first attempt to go beyond the 2+1-dimensional case in the
calculations of this type, considering the more realistic case of a
four dimensional black hole solution. We performed the calculations
considering the case of a quantized massive scalar field as a source
to the semiclassical Einstein equations. In the first section we
explain how to obtain analytical expressions for the quantum stress
tensor for a massive scalar field in the large mass limit, using a
Schwinger -DeWitt approximation to construct the one-loop effective
action, and present the results in a particular classical background
given by the Reissner-Nordstrom spacetime. Section 2 is devoted to
solve the general backreaction problem for a general spherically
symmetric spacetime, in terms of the components of the quantum
stress energy tensor, and presents the particular results obtained
for an electrically charged semiclassical black hole. In section 3
we consider the evolution of a massless test scalar field coupled to
the semiclassical background and solve the Klein-gordon equation by
separation of variables, determining the form of the effective
potential for the test field. Section 4 contains the numerical
results for the quasinormal frequencies of the semiclassical black
hole considered, and we make a comparison with the 'bare' classical
Reissner-Nordstr\"om solution. Finally in Section 5 we give the
concluding remarks and comment future related problems to be
studied.

In the following we use for the Riemann tensor, its contractions,
and the covariant derivatives the sign conventions of Misner, Thorne
and Wheeler \cite{misner}. Our units are such that
 \(\hbar=c=G=1\).

\section{Renormalized stress energy tensor for quantum scalar massive field                 }

Consider a single quantum scalar field \(\phi(x)\) with mass $m$
interacting with gravity with non minimal coupling constant \(\xi\)
in four dimensions. In the large mass limit the one loop effective
action for the quantized scalar field \cite{frolov-bv,avramidi,matyjasek1,matyjasek,owen1} is given by
\begin{eqnarray}
W_{ren}^{(1-loop)}&=&{1\over 6(4\pi)^{2}m^{2}\,} \int  d^{4}x
\sqrt{-g}\,\,str \,a_{3}(x,x),
\nonumber \\
\end{eqnarray}
where \(str a_{3}\) denotes the functional supertrace of $a_{3}$\cite{avramidi}, and \(\,a_{3}(x,x)\) is the coincidence limit
fourth Hadamard-Minakshisundaram-DeWitt-Seeley coefficient (HMDS).
As usual, the first three coefficients of the DeWitt-Schwinger
expansion, $a_{0},\,a_{1},\,{\rm and}\,a_{2}, $ contribute to the
divergent part of the action and can be absorbed in the classical
gravitational action by renormalization of the bare gravitational
and cosmological constants. Upon inserting the expression for the
HMDS coefficient in the above formula for the effective action we
obtain a renormalized effective lagrangian \cite{avramidi,matyjasek,owen1,owenarxiv} given by
\begin{equation}\label{}
    \mathfrak{L}_{ren}\,=\,\mathfrak{L}_{ren}^{conformal}+\widetilde{\mathfrak{L}_{ren}}\quad,
\end{equation}
where the conformal part of the effective lagrangian is
\begin{eqnarray}\nonumber
 \nonumber
 \mathfrak{L}_{ren}^{conformal}\,&=&\,{1\over 192 \pi^{2} m^{2}} \bigg[ \frac{1}{7560}\Theta R
 \Box R\,+\,{1\over 140} R_{\mu \nu} \Box R^{\mu \nu}-{8\over 945} R^{\mu}_{\nu} R^{\nu}_{\gamma} R^{\gamma}_{\mu}
 \\\nonumber&+&{2\over 315} R^{\mu \nu}
 R_{\gamma \varrho} R^{\gamma ~ \varrho}_{~ \mu ~ \nu}+{1\over 1260} R_{\mu \nu} R^{\mu}_{~ \sigma \gamma \varrho}
R^{\nu \sigma \gamma \varrho}+ {17\over 7560} {R_{\gamma
\varrho}}^{\mu \nu} {R_{\mu\nu}}^{\sigma \tau} {R_{ \sigma
\tau}}^{\gamma \varrho}\\ &-&{1\over 270} R^{\gamma ~ \varrho}_{~ \mu ~
\nu}
 R^{\mu ~ \nu}_{~ \sigma ~ \tau} R^{\sigma ~ \tau}_{~ \gamma ~ \varrho}\bigg]\quad ,
\end{eqnarray}
and the mass dependent contribution takes the form
\begin{equation}
 \widetilde{\mathfrak{L}_{ren}}\,=\,{1\over 192 \pi^{2} m^{2}} \left[ {1\over 30} \eta\left(R R_{\mu \nu } R^{\mu \nu}-R R_{\mu \nu \gamma \varrho} R^{\mu \nu \gamma \varrho}\right)+{1\over 2}\eta^{2} R \Box R
 \,-\,\eta^{3} R^{3}\right] ,
\end{equation}
where we use \(\Theta=30 - 252 \xi\) and \(\eta=\xi-\frac{1}{6}\).

By standard functional differentiation of the effective action with
respect to the metric,  the renormalized Stress-Energy tensor is
obtained being given by
\begin{eqnarray}\nonumber
    \langle T_{\mu\nu}\rangle_{ren}&=&\frac{2}{\sqrt{\ -g}}\frac{\delta W_{ren}}{\delta g^{\mu\nu}}\\
   &=&C_{\mu}^{\ \ \nu}+D_{\mu}^{\ \ \nu}\quad ,
   \label{emTensor}
\end{eqnarray}
where the $C_{\mu}^{\ \ \nu}$ and $D_{\mu}^{\ \ \nu}$ tensors take
cumbersome forms that the reader can find in \cite{owen1,owenarxiv}. Different but equivalent expressions for the
renormalized stress tensor at one-loop level were obtained in
references \cite{matyjasek,Folacci}.

For the present work we deal with the Reissner-Nordstr\"om
spacetime. This results can be found in the paper by Matyjasek
\cite{matyjasek1} and is amazingly simple:
\begin{eqnarray}
\langle T^{\mu}_{\nu}\rangle &=&C^{\mu}_{\nu} +\left(\xi
-\frac{1}{6}\right)D^{\mu}_{\nu}\quad,
\end{eqnarray}
where
\begin{eqnarray}\label{}\nonumber
    C^{t}_{t}&=&-\Upsilon(1248 Q^{6} -810 r^4 Q^2 +855 M^2 r^4+202 r^2
Q^4-1878 M^3 r^3 +1152 M r^3 Q^2\\\nonumber &+&2307 M^2 r^2 Q^2 -3084 M Q^4 r),
\end{eqnarray}
\begin{eqnarray}\label{}\nonumber
    D^{t}_{t}&=&\Xi(-792 M^3 r^3 +360 M^2 r^4 +2604 M^2
Q^2 r^2-1008 M Q^2 r^3 -2712 M Q^4\\\nonumber &+&819 Q^6 +728 Q^4 r^2),
\end{eqnarray}
\begin{eqnarray}\label{}\nonumber
    C^{r}_{r}&=&\Upsilon(444 Q^6 - 1488 M Q^2 r^3 +162
Q^2 r^4 +842 Q^4 r^2-1932 M Q^4 r+315 M^2 r^4\\\nonumber &+&2127 M^2 Q^2 r^2 -462
M^3 r^3),
\end{eqnarray}
\begin{eqnarray}\label{}\nonumber
    D^{r}_{r}&=&\Xi(-792 M^3 r^3 +360 M^2 r^4 +2604
M^2 Q^2 r^2 -1008 M Q^2 r^3-2712 M Q^4 r\\\nonumber &+& 819 Q^6 +728 Q^4 r^2),
\end{eqnarray}
\begin{eqnarray}\label{}\nonumber
    C^{\theta}_{\theta}&=&-\Upsilon(3044 Q^4 r^2 -2202
M^3 r^3 -10356 M Q^4 r+3066 Q^6 -4884 M Q^2 r^3\\\nonumber &+&9909 M^2 Q^2+945
M^2 r^4 +486 Q^2 r^4 ),
\end{eqnarray}
\begin{eqnarray}\label{}\nonumber
    D^{\theta}_{\theta}&=&\Xi(3276
M^2 Q^2 r^2-1176 M Q^2 r^3 -3408 M Q^4 r +1053 Q^6-1008 M^3 r^3 \\\nonumber&+&
432 M^2 r^4 +832 Q^4 r^2),
\end{eqnarray}
In the above expressions, we have \(\Upsilon=\left(30240 \pi^2 m^2
r^{12}\right)^{-1}\) and \(\Xi=\left(720 \pi^2
m^2r^12\right)^{-1}\). $Q$ and $M$ denotes the charge and bare mass
of the black hole.

These results were obtained using the expression for the stress
energy tensor presented in \cite{owen1} and coincides with that
previously obtained by Matyjasek in \cite{matyjasek1} using also the
Schwinger-DeWitt approximation for Ricci flat spacetimes and by
Anderson, Hiscock and Samuel in reference \cite{AHS} using sixth
order WKB approximation for the mode functions of the Klein-Gordon
equation in the background spacetime. Moreover, the above stress
energy tensor is covariantly conserved, and in the limit of zero
black hole electric charge we obtain the results of Frolov and
Zelnikov for the Schwarzshild spacetime. Also the quantum stress
energy tensor is regular at the event horizon, as is to be expected
due to the local nature of the Schwinger-DeWitt approximation and
the regular nature of the horizon.

In our chosen system of units the general condition for the validity
of the Schwinger-DeWitt approximation can be put as \(mM\geq1\),
where \(m\) and \(M\) are respectively the scalar field and black
hole masses. A more specific condition valid for Reissner-Nordstrom
( R-N ) spacetimes was given by Anderson et.al in Reference
\cite{AHS}, where they show, using detailed numerical results, that
for \(mM\geq2\) the deviation of the approximate stress energy
tensor from the exact one lies within a few percent. In all the
numerical calculations presented in the following sections of this
paper, we carefully take into account the fulfillment of this
condition.

In the next section we use these results to get the solution of
semiclassical Einstein equations in the form of a semiclassical
electrically charged black hole.

\section{Semiclassical solution}

In this section we show how to solve the general backreaction
problem for spherically symmetric spacetimes applying general
results to the particular case of an electrically charged
semiclassical black hole, obtaining the quantum correction to the
classical Reissner-Nordstr\"om metric. There are previous studies on
this direction \cite{berej-matyjasek,taylor}, but we present here
general results that can be applied to any spherically symmetric
spacetime, following the lines of reference \cite{lousto-sanchez}.
Consider the line element for a general spherically symmetric
spacetime
\begin{equation}\label{ssmetric}
     ds^{2}=-A(r)dt^{2}+B(r)dr^{2}+r^{2}\left(d\theta^{2}+\sin^{2}\theta
     d\phi^{2}\right),
\end{equation}
We intend to solve the general semiclassical Einstein equations with
the source \(T_{\mu\nu}=T_{\mu\nu}^{class}+\left\langle
T_{\mu\nu}\right\rangle\) including two contributions: the first,
denoted by \(T_{\mu\nu}^{class}\), comes from a classical source,
and the second, denoted by \(\left\langle T_{\mu\nu}\right\rangle\),
is the quantum field contribution. In the following, we assume that
the classical source is an electromagnetic field, so the solution to
the backreaction problem gives a quantum corrected
Reissner-N\"{o}rstrom black hole. It is possible to show, using
appropiate combinations of the components of the Ricci tensor for
the line element (\ref{ssmetric}), that the general form for the
metric components \(g_{rr}=B(r)\) and \(-g_{tt}=A(r)\) that solves
the backreaction problem are given by
\begin{equation}\label{grrcomponent}
    \frac{1}{B(r)}=1-\frac{2M}{r}+\frac{Q^{2}}{r^{2}}+\frac{8 \pi}{r} \int_{\infty}^{r} \zeta^{2} \left\langle\ T_{t}^{t}
    \right\rangle d\zeta ,
\end{equation}
and
\begin{equation}\label{gttcomponent1}
    A(r)=\frac{1}{B(r)}\exp \left\{\lambda(r)\right\},
\end{equation}
where
\begin{equation}\label{gttcomponent2}
    \lambda(r)=8 \pi \int_{\infty}^{r} \zeta B\left(\zeta\right)\left(\left\langle\ T_{r}^{r}
    \right\rangle- \left\langle\ T_{t}^{t}
    \right\rangle\right)d\zeta .
\end{equation}
In the above equations \(Q\) and \(M\) denotes the charge and the
bare mass of the black hole, i.e,  of the classical
Reissner-N\"{o}rdstr\"om solution. As we can easily see from
expression (\ref{grrcomponent}) the mass of the black hole changes
due to quantum effects. Note that the general solutions above are
obtained using the same boundary conditions of reference
\cite{lousto-sanchez}, which differs from that of Berej and
Matyjasek in reference \cite{berej-matyjasek} that uses a horizon
defined mass as the parameter in their final expressions. Usually,
the integrals in (\ref{grrcomponent}) and (\ref{gttcomponent2}) are
performed introducing some perturbation approximation due to the
fact that the quantum stress tensor \(\left\langle
T_{\mu\nu}\right\rangle\) is linear in the Dirac constant \(\hbar\)
( that in our chosen units is \(\hbar=1\) ). In this sense, we are
following the idea mentioned in the introduction, consistent in
considering the right hand side of the semiclassical Einstein
equations as a perturbation, since may be, therefore, the only one
way to obtain the approximate physical solution to the backreaction
problem. We use as a perturbation parameter the ratio
\(\varepsilon=1/M^{2}\), where $M$ is the mass of the black hole (
in conventional units we have \(\varepsilon=M_{P}^{2}/M^{2}\), where
\(M_{P}\) is the Planck's mass). Now inserting in
(\ref{grrcomponent}) and (\ref{gttcomponent2}) the general
expressions for the quantum stress tensor evaluated in the classical
Reissner-Nordstr\"om metric and considering only terms that are
linear in the perturbation parameter we can obtain \(A(r)\) and
\(B(r)\) up to \(\mathcal{O}(\varepsilon^{2})\). After some algebra
we obtain
\begin{equation}\label{grrcomponentfinal}
    \frac{1}{B(r)}=1-\frac{2M}{r}+\frac{Q^{2}}{r^{2}}+\frac{\varepsilon}{\pi m^{2}}\left(F(r)+\xi
    H(r)\right),
\end{equation}
where
\begin{eqnarray}\label{f}\nonumber
F(r)&=&-\frac{613 M^{3}Q^4}{840 r^9} +\frac{2327 M^{2}Q^{6}}{1134
r^{10}} -\frac{3 M^{2}Q^2}{70 r^6} +\frac{5 M^{4}}{28 r^6}
-\frac{1237 M^{5} }{3780 r^{7}} +\frac{883 M^{2}Q^4}{4410
r^{8}}\\\nonumber&-&\frac{82 M^{3}Q^{^2}}{315 r^7}+\frac{1369
M^{4}Q^{2}}{1764r^8},
\end{eqnarray}
\begin{equation}\label{h}\nonumber
H(r)=\frac{28 M^{3}Q^{2}}{15 r^7}+\frac{113M^{3}Q^{4}}{30 r^{9}}
-\frac{91 M^{2}Q^{6}}{90 r^10}-\frac{52 M^{2}Q^{4}}{45 r^{8}}-
\frac{4M^{4}}{5 r^{6}} +\frac{22 M^{5}}{15 r^{7}} -\frac{62
M^{4}Q^{2}}{15 r^{8}}.
\end{equation}

For the function \(\lambda(r)\) in (\ref{gttcomponent1}) defined by
(\ref{gttcomponent2}) we obtain
\begin{eqnarray}\label{lambda}\nonumber
\lambda(r)&=&\frac{\varepsilon }{\pi m^2}\left(\frac{184M^{3}Q^{2}
}{441 r^{7}}-\frac{29 M^{4}}{140 r^{6}}-\frac{229 M^{2}Q^{4}}{840
r^{8}}+\frac{M^{2}Q^{2}}{35 r^{6}}\right)\\\nonumber&+&\frac{\varepsilon  \xi
}{\pi m^2}\left(\frac{14M^{4}}{15 r^{6}}+\frac{13 M^{2}Q^{4}}{10
r^{8}} -\frac{32 M^{3} Q^{2}}{15 r^{7}} \right).
\end{eqnarray}
With the above analytical results for the semiclassical line element
representing a quantum corrected charged black hole, we can
determine the changes in some of the properties of the solution,
with respect to its classical counterpart. In particular, we can
determine the change in the position of the event horizon due to
quantum effects. Let be \(r_{h}=M+\sqrt{M^{2}-Q^{2}}\) the position
of the event horizon for the classical charged black hole. The
horizon for the quantum corrected solution will be at position
\(r_{+}\) defined by
\begin{equation}\label{horizoneqn}
    A(r_{+})=\frac{1}{B(r_{+})}=1-\frac{2M}{r_{+}}+\frac{Q^{2}}{r_{+}^{2}}+\frac{8 \pi}{r_{+}} \int_{\infty}^{r_{+}} \zeta^{2} \left\langle\ T_{t}^{t}
    \right\rangle d\zeta=0 .
\end{equation}

Solving (\ref{horizoneqn}) we find that, up to first order in the
perturbation parameter, the exact horizon position \(r_{+}\) for the
semiclassical charged black hole is given by
\begin{equation}\label{exacthorizon}
    r_{+}=r_{h}\left(1+\varepsilon \Lambda\right),
\end{equation}
where
\begin{equation}\label{shifthorizon1}
    \Lambda=-\frac{4\pi}{\left(M-Q^{2}/r_{h}\right)}\int_{\infty}^{r_{h}}\zeta^{2}\left\langle T_{t}^{t}(\zeta)\right\rangle\
    d\zeta .
\end{equation}
Now using the expressions for the temporal component of the
stress-energy tensor of the quantum field, and performing the above
integral, we obtain as the final result
\begin{equation}\label{shifthorizonresult}
    \Lambda=\frac{\varepsilon}{\pi
    m^{2}\left(M-Q^{2}/r_{h}\right)}\left(\Gamma+\xi\Omega\right),
\end{equation}
with
\begin{eqnarray}\label{gamma}\nonumber
\Gamma&=&\frac{613 M^{3}Q^4}{1680 r_{h}^8} -\frac{2327
M^{2}Q^{6}}{22680 r_{h}^{9}} +\frac{3 M^{2}Q^2}{140 r_{h}^5}
-\frac{5 M^{4}}{56 r_{h}^6} +\frac{1237 M^{5} }{7560 r_{h}^{6}}
-\frac{883 M^{2}Q^4}{8820 r_{h}^{7}}\\\nonumber&+&\frac{41 M^{3}Q^{^2}}{315
r_{h}^6}-\frac{1369 M^{4}Q^{2}}{3528 r_{h}^7},
\end{eqnarray}
\begin{equation}\label{h}\nonumber
\Omega=-\frac{14 M^{3}Q^{2}}{15 r_{h}^6}-\frac{113M^{3}Q^{4}}{60
r_{h}^{8}} +\frac{91 M^{2}Q^{6}}{180 r_{h}^9}+\frac{26
M^{2}Q^{4}}{45 r_{h}^{7}}+ \frac{2M^{4}}{5 r_{h}^{5}} -\frac{11
M^{5}}{15 r_{h}^{6}} +\frac{31 M^{4}Q^{2}}{15 r_{h}^{7}}.
\end{equation}
It is interesting to note that the effect of the quantum field over
the bare black hole spacetime is to reduce the position of the event
horizon. This reduction is a consequence of the typical fact that
the weak energy condition for the quantum field is violated on the
event horizon.

\section{Looking for scalar quasinormal frequencies}

In this section, we consider the evolution of a test massless scalar
field $\Phi(x^{\mu})$, where $x^{\mu}=(t,r,\theta,\phi)$, in the
background of the semiclassical spherically charged black hole
studied above. The dynamics of  for this test field is governed by
the Klein-Gordon equation
\begin{equation}\label{kg1}
\frac{1}{\sqrt{-g}}\frac{\partial}{\partial
x^{\mu}}\left(\sqrt{-g}g^{\mu\nu}\frac{\partial\Phi}{\partial
x^{\nu}}\right)=0\quad,
\end{equation}
with $g_{\mu\nu}$ is the metric tensor of semiclassical solution
(\ref{grrcomponent}) to (\ref{lambda}), and $g$ its determinant.

Changing the wave function $\Phi=\Psi/r$, and the radial coordinate
$dr/dr_{*}=\sqrt{B(r)/A(r)}$, and separating the time, radial and
angular dependence of the field as $\Psi=e^{i\omega
t}Z(r)_{L}Y_{Lm}(\theta,\phi)$, the Klein-Gordon equation is written
as
\begin{equation}\label{kg2}
\frac{d^2}{dr^{2}_{*}}Z_{L}+\left[\omega^2 - V\right]Z_{L}=0\quad,
\end{equation}
where $\omega$ is the quasinormal frequency and $V$ is the effective
potential. The potential $V$ is a function of the metric components
$g_{\mu\nu}$ and the multipolar number $L$,
\begin{equation}\label{potencial}
V[r(r_{*})]=A(r)\frac{L(L+1)}{r^2}+\frac{A(r)}{2rB(r)}\left[\left(\ln{A(r)}\right)'
- \left(\ln{B(r)}\right)'\right]\quad,
\end{equation}
where the primes refer to the derivatives with respect to the radial
coordinate $r$. For the specific case of semiclassical electrically
charged black holes discussed in this paper, we have for the
effective potential the general result
\begin{equation}\label{}
    V(r)=V^{c}(r)+\frac{\varepsilon}{\pi \,{m}^ {2}}U(r) +O \left( {\epsilon}^{2}
    \right),
\end{equation}
where \(V^{c}(r)\) is the scalar effective potential of the bare
Reissner-Nordstrom black hole given by
\begin{equation}\label{barepotential}
    V^{c}(r)={\frac { \left( {r}^{2}-2\,Mr+{Q}^{2} \right)  \left( -2\,{Q}^{2}+
\beta\,{r}^{2}+2\,Mr \right) }{{r}^{6}}},
\end{equation}
and \(U(r)\) is proportional to the first order contribution of the
vacuum polarization effect to the total effective potential.
The
general expression for this magnitude can be written as
\begin{equation}\label{}
    U(r)=W_{1}(r)+\xi W_{2}(r),
\end{equation}
where
\begin{eqnarray}\nonumber
W_{1}(r)=&-&{\frac {1751}{4410}}\,{\frac
{{M}^{2}{Q}^{4}}{{r}^{10}}}-{\frac {9}{20}}\,{\frac
{{M}^{4}}{{r}^{8}}}+ {\frac {1021}{540}}\,{\frac
{{M}^{5}}{{r}^{9}}}-{\frac { 1816}{945}}\,{\frac
{{M}^{6}}{{r}^{10}}}+{\frac {6}{35}} \,{\frac
{{M}^{2}{Q}^{2}}{{r}^{8}}}\\\nonumber&+&{\frac {674641}{ 158760}}\,{\frac
{{M}^{3}{Q}^{6}}{{r}^{13}}}+{\frac {17}{ 105}}\,{\frac
{{M}^{3}{Q}^{2}}{{r}^{9}}}
         -{\frac {13271}{1764}}\,{ \frac
{{M}^{4}{Q}^{4}}{{r}^{12}}}-{\frac {625}{756}}\,{\frac {{M}^{2}{
Q}^{8}}{{r}^{14}}}\\\nonumber&-&{\frac {23353}{15876}} \,{\frac
{{M}^{2}{Q}^{6}}{{r}^{12}}}+{\frac {8559}{1960}} \,{\frac
{{M}^{3}{Q}^{4}}{{r}^{11}}}-{\frac {962}{245}}\,{\frac {{M}^
{4}{Q}^{2}}{{r}^{10}}}+{\frac {16687}{2940}}\,{\frac {{M}^{5}
{Q}^{2}}{{r}^{11}}}
         \\\nonumber &+& L\left(L+1\right)\bigg(-{\frac {1529}{22680}}\,{\frac
{{M}^ {2}{Q}^{6}}{{r}^{12}}}-\frac{1}{35}\,{\frac {{M
}^{4}}{{r}^{8}}}-{\frac {1}{70}}\,{\frac {{M}^
{2}{Q}^{2}}{{r}^{8}}}\\\nonumber &+&{\frac {47}{540}}\,{\frac {{M}
^{5}}{{r}^{9}}}-{\frac {773}{17640}}\,{\frac {{M}^{2}
{Q}^{4}}{{r}^{10}}}+{\frac {44}{441}}\,{\frac {{M}^{3}
{Q}^{2}}{{r}^{9}}}+{\frac
{821}{3528}}\,{\frac {{M}^ {3}{Q}^{4}}{{r}^{11}}}-{\frac
{1171}{4410}}\,{\frac {{M}^{4}{Q}^{2}}{{r}^{10}}}\bigg),
\end{eqnarray}
and
\begin{eqnarray}\nonumber
W_{2}(r)=&&L\left(L+1\right)\bigg(-{\frac {4}{15}}\,{\frac
{{M}^{3}{Q}^{2}}{{r}^{9}}}+ {\frac {16}{15}}\,{\frac
{{M}^{4}{Q}^{2}}{{r}^{10}}} -{\frac {29}{30}} \,{\frac
{{M}^{3}{Q}^{4}}{{r}^{11}}}-{\frac {2}{5}}{\frac
{{M}^{5}}{{r}^{9}}}+{\frac {13}{90}}\,{\frac
{{M}^{2}{Q}^{4}}{{r}^{10}}}\\\nonumber&+&{\frac {13}{45}}\,{\frac
{{M}^{2}{Q}^{6}}{{r}^{12}}}+\frac{2}{15}\,{ \frac
{{M}^{4}}{{r}^{8}}}\bigg)
+ {\frac {26}{3}}\,{\frac
{{M}^{2}{Q}^{6}}{{r}^ {12}}}-{\frac {162}{5}}\,{\frac
{{M}^{5}{Q}^{2}}{{r}^{11} }}-{\frac {2101}{90}}\,{\frac
{{M}^{3}{Q}^{6}}{{r}^{13}}}\\\nonumber &+&{\frac {128}{15}}\,{\frac
{{M}^{6}}{{r}^{10}}}+2\,{\frac
{{M}^{4}}{{r}^{8}}}+\frac{13}{3}\,{\frac
{{M}^{2}{Q}^{8}}{{r}^{14}}}+{\frac {182}{ 45}}\,{\frac
{{M}^{2}{Q}^{4}}{{r}^{10}}}-{\frac {289}{10 }}\,{\frac
{{M}^{3}{Q}^{4}}{{r}^{11}}}-{\frac {28}{5}}\,{ \frac
{{M}^{3}{Q}^{2}}{{r}^{9}}}\\\nonumber &-&{\frac {42}{5}}\,{\frac
{{M}^{5}}{{r}^{9}}}+{\frac {416}{15}}\,{\frac
{{M}^{4}{Q}^{2}}{{r}^{10}}}+{\frac {130}{3}}\,{\frac
{{M}^{4}{Q}^{4}}{{r}^{12}}}.
\end{eqnarray}

\begin{figure}[htb!] 
           \includegraphics[width=11cm]{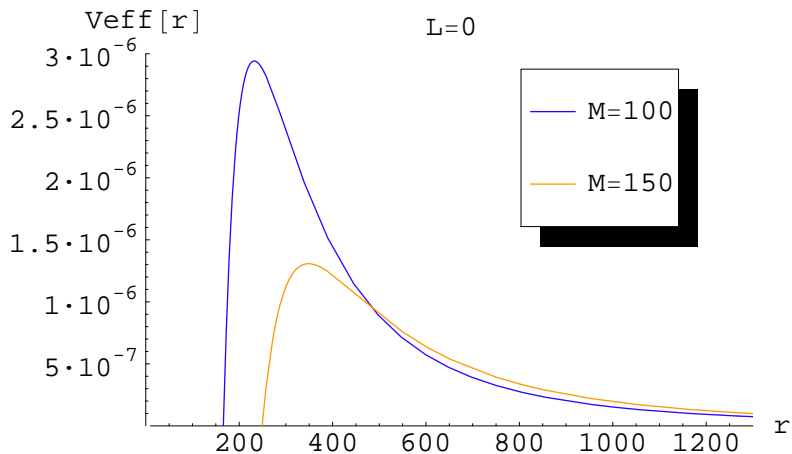}\\
           \caption{\it Effective potential of L=0 scalar modes for Semiclassical black hole with $M=100$(top curve) and $M=150$(bottom curve). $Q/M=0.75$ and the mass parameter of the quantum scalar field is chosen to be m=1/10. }
           \label{potencial}
       \end{figure}

As we can see in diagram (\ref{potencial}), the effective potential
$V$ has the form of a definite positive potential barrier, i.e, it
is a well behaved function that goes to zero at spatial infinity and
gets a maximum value near the event horizon. The quasinormal modes
are solutions of the wave equation (\ref{kg2}) with the specific
boundary conditions requiring pure out-going waves at spatial
infinity and pure in-coming waves on the event horizon. Thus no
waves come from infinity or the event horizon. The quasinormal
frequencies are in general complex numbers, whose real part
determines the real oscillation frequency and the imaginary part
determines the damping rate of the quasinormal mode. In order to
evaluate the quasinormal modes for the test scalar field, we use the
well known WKB technique, that can give accurate values of the
lowest ( that is longer lived ) quasinormal frequencies. The first
order WKB technique was applied to finding quasinormal modes for the
first time by Shutz and Will \cite{shutz-will}. Latter this approach
was extended to the third order beyond the eikonal approximation by
Iyer and Will \cite{iyer-will} and to the sixth order by Konoplya
\cite{konoplya2}.

We use in our numerical calculation of quasinormal modes this sixth
order WKB expansion, for which was shown in \cite{konoplya3} that
gives a relative error which is about two order less than the third
WKB order. The sixth order WKB formula can be written as
\begin{equation}\label{WKB6}
    i\frac{\left(\omega^{2}-V_{0}\right)}{\sqrt{-2V_{0}^{''}}}-\sum_{j=2}^{6}\Pi_{j}=n+\frac{1}{2},
    \ \ \ \ \ n=0,1,2,...
\end{equation}
where \(V_{0}\) is the value of the potential at its maximum as a
function of the tortoise coordinate, and \(V_{0}^{''}\) represents
the second derivative of the potential with respect to the tortoise
coordinate at its peak. The correction terms \(\Pi_{j}\) depend on
the value of the effective potential and its derivatives ( up to the
2i-th order) in the maximum ( see \cite{zhidenkothesis} and
references therein ). The above formula was used in several papers
for the determination of quasinormal frequencies in a variety of
systems \cite{WKB6papers}. Yet one should remember that strictly
speaking the WKB series converge only asymptotically and the
consequence decreasing of the relative error in each WKB order is
not guaranteed. Fortunately, in the considered case here, WKB series
shown convergence in all sixth orders.

In the following we show the results for the numerical evaluation of
the first two fundamental quasinormal frequencies considering a
minimal coupling between the quantum field and the bare black hole.
As we can see for the above expressions for the effective potential
of the semiclassical black hole, the parameters entering in the
determination of the quasinormal frequencies are the black hole bare
charge \(Q\) and mass \(M\), the coupling constant \(\xi\) between
the quantum scalar and the classical gravitational fields, and the
quantum field mass \(m\). In table (1) we list values for the real
and imaginary parts of the quasinormal frequencies for semiclassical
and classical black holes varying the multipolar number $L$ and the
overtone number $n$. The numerical results are obtained for minimal
coupling between the quantum field and the background gravity field,
fixing the bare black hole charge to mass ratio, and allowing to
variation in \(M\). The quantum scalar field mass is fixed taking
into account the condition for the validity of the Schwinger-DeWitt
approach.

\begin{table}[htb!]\label{frequencias_quantico}
   \begin{center}
   \setlength{\arrayrulewidth}{2\arrayrulewidth}  
   \setlength{\belowcaptionskip}{5pt}  
   \begin{tabular}{|c|c|c|c|c|c|c|c|}
      \hline
      \multicolumn{4}{||c|}{Semiclassical solution}&\multicolumn{4}{|c||}{Classical solution} \\
      \hline
      \hline
      \multicolumn{8}{|c|}{$M=100$} \\
      \hline
      $L$ & $n$ & $Re(\varpi)$ & $-Im(\varpi)$ & $L$ & $n$ &
      $Re(\varpi)$  &$-Im(\varpi)$ \\
      \hline
  $ 0 $ & $0$& 2.3302  & 1.2140 &$ 0$ & $ 0 $  & 14.233  &1.0638  \\
      \hline
  $ 1 $& $0$ & 3.5226 & 1.1772  &$ 1 $&$ 0$    & 3.8461  &  1.1606 \\
      \hline
 $ 1 $& $1$ & 3.1836   & 3.6897    &$1 $ &$ 1$    & 8.1164 & 2.4344 \\
     \hline
       \hline
      \multicolumn{8}{|c|}{$M=110$}  \\
      \hline
      $L$ & $n$ & $Re(\varpi)$ & $-Im(\varpi)$ & $L$ & $n$ &
      $Re(\varpi)$  &$-Im(\varpi)$ \\
      \hline
  $ 0 $ & $0$& 1.2090 & 1.1042 &$ 0$ & $ 0 $  & 6.0223  & 0.9657 \\
      \hline
  $ 1 $& $0$ & 3.2054 &  1.0698  &$ 1 $&$ 0$    & 3.4741   &  0.9579\\
      \hline
 $ 1 $& $1$ & 2.8942   & 3.3544   &$1 $ &$ 1$    & 5.1130 & 1.5760 \\
     \hline
       \hline
      \multicolumn{8}{|c|}{$M=150$}  \\
      \hline
      $L$ & $n$ & $Re(\varpi)$ & $-Im(\varpi)$ & $L$ & $n$ &
      $Re(\varpi)$  &$-Im(\varpi)$ \\
      \hline
  $ 0 $ & $0$& 1.0230 & 0.9337 &$ 0$ & $ 0 $  & 2.1751  & 0.8333 \\
      \hline
  $ 1 $& $0$ & 2.7123  & 0.9053   &$ 1 $&$ 0$    &2.9588   & 0.8928 \\
      \hline
 $ 1 $& $1$ & 2.4490   & 2.8383     &$1 $ &$ 1$    &6.2429  & 1.8724 \\
     \hline
       \hline
   \end{tabular}
   \caption{\it Rescaled Scalar quasinormal frequencies $\varpi=10^{3}\omega$ for the classical and semiclassical Reissner-Nordstr\"om black hole, with
   $Q/M=0.75$, $m=1/10$ and $\xi=0$.}
   \end{center}
  \end{table}

\begin{figure}[htb]
       \includegraphics[width=9cm]{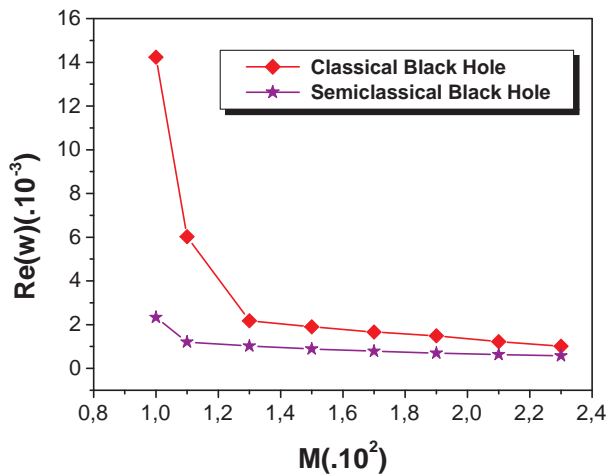}\\
           \caption{\it Dependance of $Re(\omega)$ on M for classical and semiclassical black holes. The parameters are chosen to be $Q/M=0.75$, $m=1/10$, $\xi=0$, $L=0$ and $n=0$.}
           \label{comparacao1}
\end{figure}

\begin{figure}[htb]
       \includegraphics[width=9cm]{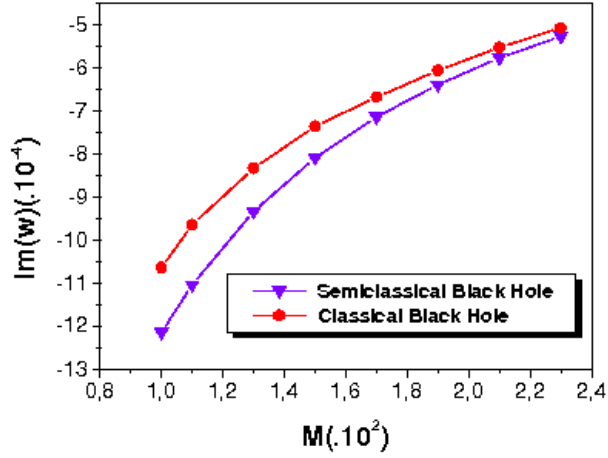}\\
           \caption{\it Dependance of $Im(\omega)$ on M for classical and semiclassical black holes. The parameters are chosen to be $Q/M=0.75$, $m=1/10$, $\xi=0$, $L=0$ and $n=0$.}
           \label{comparacao2}
\end{figure}

As the obtained numerical results show, a shift in the quasinormal
spectrum due to semiclassical corrections of the
Reissner-Nordstr\"om background appears, an effect that is more
pronounced for the fundamental mode \(( L=0 )\). From
(\ref{comparacao1}) and (\ref{comparacao2}) we see that the
backreaction of the quantized field upon the classical solution
gives rise to a decreasing of the real oscillation frequencies and
to a small decreasing of the damping rate, for physically
interesting values of the black hole mass. Using the numerical
results above presented, it is simple to see that, as a consequence
of the vacuum polarization effect, we have an effective decreasing
of the quality factor, proportional to the ratio
\(\frac{\left|Re(\omega)\right|}{\left|Im(\omega)\right|}\). As
expected, the differences in the quasinormal frequencies when the
black hole mass increases tend to disappear.
From the above results, we arrive to the conclusion that the
classical Reissner-Nordstr\"om black holes are better oscillators
than its quantum corrected partners. This is in contrast with the
results obtained by Konoplya in reference \cite{konoplyabtz} for the
BTZ black hole dressed by a quantum conformal massless scalar field.
In that case was shown that the backreaction of the Hawking
radiation increases the quality factor for semiclassical BTZ black
holes in the small mass regime. The investigation of a similar
problem (i.e, the increasing of the quality factor due to Hawking
radiation, in contrast with the decreasing of that magnitude due to
vacuum polarization ) for semiclassical charged black holes is in
progress, and will be the subject of future publication by us.

\begin{figure}[htb]
           \includegraphics[width=9cm]{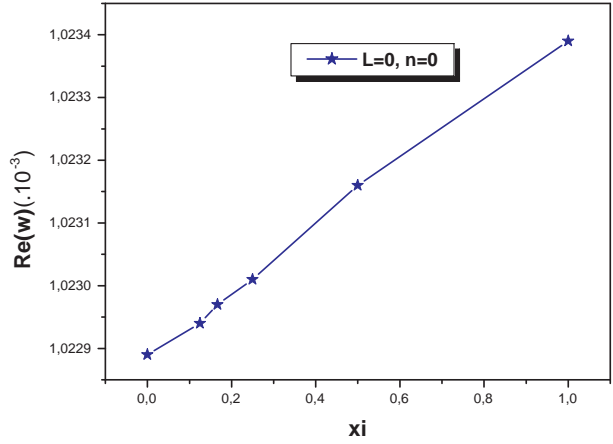}\\
           \caption{\it Behaviour of the real part of the quasinormal frequencies of the fundamental mode for different values of couplig constant $\xi$. In the figure $M=100$, $Q/M=0.75$,$m=1/10$.}
           \label{acoplamento1}
\end{figure}

\begin{figure}[htb]
           \includegraphics[width=9cm]{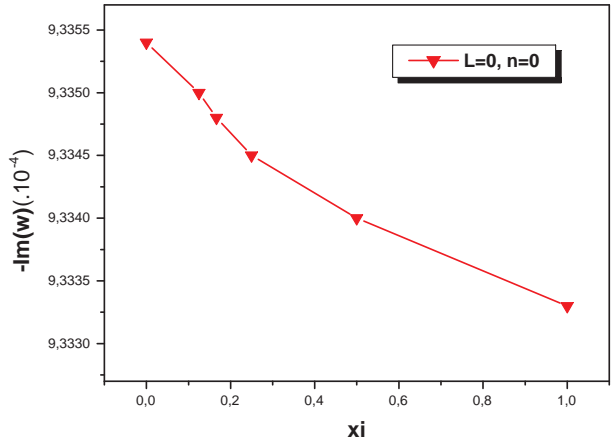}\\
           \caption{\it Behaviour of the imaginary part of the quasinormal frequencies of the fundamental mode for different values of couplig constant $\xi$. In the figure $M=100$, $Q/M=0.75$,$m=1/10$.}
           \label{acoplamento2}
\end{figure}
We also evaluate the dependence of the quasinormal frequencies for a
given black hole bare mass and different values of the coupling
constant between the quantum field of fixed mass $m$ and the
classical background spacetime, including the more physically
interesting cases of minimal and conformal coupling . The results
appears in figure (\ref{acoplamento1}) and (\ref{acoplamento2}). As
we can see, the quasinormal frequencies shows only a little
dependance on the coupling constant. As the coupling constant
increases, we see an almost linear small increment in the real and
imaginary parts of the quasinormal frequencies for semiclassical
black holes. A similar very small effect appears if we consider the
dependance of the quasinormal frequencies on the mass of the quantum
scalar field, for a fixed bare black hole mass and coupling
constant. As the quantum field mass increases, we found a very
little increment in the real part of the frequencies for
semiclassical black holes, and a very little decreasing in the
imaginary part. Therefore, the shift in the quasinormal frequencies
with respect to the classical bare black hole appears to be almost
the same for the given range of the quantum field parameters.

\section{Concluding remarks}

We have studied the influence of the backreaction due to vacuum
polarization on the structure of test scalar quasinormal frequencies
for semiclassical charged black holes. The semiclasical solution
studied describes a Reissner-Nordstr\"om black hole dressed by a
quantum massive scalar field in the large mass limit. Such an
influence appears essentially as an appreciable shift in the
quasinormal frequencies that decreases as the bare black hole mass
increases, and that not have a strong dependance upon the quantum
field parameters. This shift shows that the quantum corrected
quasinormal modes are less oscillatory with respect to its classical
counterpart. The new features mentioned above regarding the
quasinormal frequencies of the semiclassical solution so far reflect
the physics of the back reaction of a single specie of a field of
spin $s=0$. While this have revealed novel important features of the
quasinormal ringing stage for the quantum corrected black hole
considered, a more realistic setting should take into account black
holes surrounded by the multiple species of quantum fields,
belonging to the Standard Model. It is expected that the shifts in
the frequency spectrum obtained in this case should be more
appreciable than in the situation considered in this work. The
results for this more realistic case will be presented in a
forthcoming paper.

\section{Acknowledgements}
This work has been supported by FAPESP and CNPQ, Brazil as well as
ICTP, Trieste. We are grateful to Professor Elcio Abdalla for the
useful suggestions. One of the authors (J. de Oliveira) ought to
express his gratitude by the kind hospitality granted by the
Departamento de F\'{i}sica y Qu\'{i}mica at the University of
Cienfuegos, Cuba, where this work was realized. O.P. F. Piedra also
thanks to Owen Daniel Fern\'{a}ndez Chac\'{o}n, for his help in
preparing the figures, and to Dr. Alexander Zhidenko, for provide
valuable information about numerical methods usually employed in the
calculation of quasinormal frequencies.


\end{document}